\documentclass[12pt]{article}
\usepackage{amstex}
\usepackage{amssymb}
\usepackage{a4}
\usepackage{epsfig}
\begin{document}
\newcommand{\newc}{\newcommand}
\newc{\R}{$R$}
\newc{\charginom}{M_{\tilde \chi}^{+}}
\newc{\mue}{\mu_{\tilde{e}_{iL}}}
\newc{\mud}{\mu_{\tilde{d}_{jL}}}
\newc{\barr}{\begin{eqnarray}}
\newc{\earr}{\end{eqnarray}}
\newc{\beq}{\begin{equation}}
\newc{\eeq}{\end{equation}}
\newc{\ra}{\rightarrow}
\newc{\lam}{\lambda}
\newc{\eps}{\epsilon}
\newc{\gev}{\,GeV}
\newc{\tev}{\,TeV}
\newc{\eq}[1]{(\ref{eq:#1})}
\newc{\eqs}[2]{(\ref{eq:#1},\ref{eq:#2})}
\newc{\etal}{{\it et al.}\ }
\newc{\eg}{{\it e.g.}\ }
\newc{\ie}{{\it i.e.}\ }
\newc{\nonum}{\nonumber}
\newc{\lab}[1]{\label{eq:#1}}
\newc{\dpr}[2]{({#1}\cdot{#2})}
\newc{\gsim}{\stackrel{>}{\sim}}
\newc{\lsim}{\stackrel{<}{\sim}}

\begin{titlepage}
\begin{flushright}
{ETHZ-IPP  PR-96-02} \\
{RAL-96-049} \\ \today \\
\end{flushright}

\begin{center}
{\bf \Large
\bf How to find a Higgs Boson with a Mass between 
155--180 GeV at the LHC}
\end{center}
\smallskip \smallskip \bigskip
\begin{center}
{\Large M. Dittmar$^1$, H. Dreiner$^2$}
\end{center}
\bigskip

\begin{center}
$^1$Institute for Particle Physics (IPP), ETH Z\"{u}rich, \\
CH-8093 Z\"{u}rich, Switzerland
\end{center}

\begin{center}
$^2$ Rutherford Laboratory, Chilton, Didcot, OX11 0QX, UK
\end{center}

\begin{abstract}
\noindent We reconsider the signature of events with 
two charged leptons and
missing energy as a signal for the detection of the 
Standard Model Higgs
boson in the mass region $M(Higgs)$=155--180 GeV. 
It is shown that a few
simple experimental criteria allow to distinguish 
events originating from the
Higgs boson decaying to $H\ra W^+W^-$ from the non 
resonant production of $W^
+W^- X$ at the LHC. With this set of cuts, signal 
to background ratios of
about one to one are obtained, allowing a 5--10 $\sigma$ 
detection with about
5~$fb^{-1}$ of luminosity. This corresponds to less than 
one year of running
at the initial lower luminosity ${\cal L}= 10^{33}cm^{-2}s^{-1}$. 
This is
significantly better than for the hitherto considered Higgs 
detection mode
$H\ra Z^0Z^{0*}\ra 2\ell^+2\ell^-$, where in this mass range 
about 100~$fb^{-1}$ of
integrated luminosity are required for a 5 $\sigma$ 
signal.
\vspace{2cm}
\end{abstract}

\begin{center}
{\it Submitted to Physical Review D }
\end{center}
\end{titlepage}

\section{Introduction}
The Standard Model of elementary particle physics (SM) has 
been highly
successful in explaining all experimental data \cite{sm}. 
With the recent
discovery of the top quark \cite{cdf}, the Higgs boson is 
the only remaining
missing piece, albeit an essential one. Within the SM, 
it provides the
mechanism to dynamically break the electroweak 
symmetry {\it and} gives
masses
to the electroweak gauge bosons. The same mechanism 
gives masses to the spin
1/2 fermions such as the top quark and the electron. 
The Higgs is therefore
essential for our understanding of mass. Furthermore, 
the theoretical
structure of the Higgs sector in the SM is the main motivation for
speculations on physics beyond the SM, \eg supersymmetry
or technicolour. The discovery of the Higgs boson and the 
determination of
its couplings could possibly provide an essential clue to 
this new realm. The
search for the Higgs boson is therefore considered to be {\it the} 
most
important task for future collider physics.

With the present energy upgrade of the large electron positron 
collider, LEP2,
at CERN, the Higgs boson can be discovered for $M_{H} \leq 98$ GeV
\cite{lep2rep} ($\sqrt{s}=192$ GeV). For larger masses the 
Large Hadron
Collider (LHC) to be built at CERN, is the most promising 
discovery machine.
There, the Higgs boson search is usually split into three Higgs 
mass regions
\cite{froidevaux_lhc,cms,atlas}
\barr
&(i)\quad 90\rm{GeV} &< \,\;M_{H} \,\;<\,\; 130\, \rm{~GeV},    
   \nonum \\
&(ii)\quad 130\rm{GeV}  &< \,\;M_{H} \,\;<\,\; 2\cdot M_{Z^0},   \\
&(iii) \quad 2\cdot M_{Z^0} &< \,\;M_{H} \;\,<\;\, 800\,\rm{~GeV}. 
\nonum
\earr
For the mass regions $(i)$ and $(iii)$, Higgs detection with 
large significance
is possible by the observation of narrow mass peaks using the 
decays $H\ra
\gamma \gamma$ and $H\ra Z^{0}Z^{0}\ra 2
\ell^+2\ell^-$ respectively 
\cite{froidevaux_lhc,cms,atlas}.
For most of the
mass region $(ii)$ previous experimental studies, assuming 
excellent
energy and momentum measurements of electrons and muons, 
have obtained
promising mass peaks from the channel 
$H\ra Z^{0}Z^{*0}\ra2\ell^+2\ell^-$,
despite the low branching ratios 
\cite{froidevaux_lhc,cms,atlas}. 
However, the
mass range between $\approx$ 155--180 GeV remains 1 
difficult to
detect because the Higgs decays almost exclusively to a pair 
of on shell
$W^{\pm}$'s \cite{froidevaux_lhc,cms,atlas,lhczoltan,hunter}. 
Consequently, a
large integrated luminosity of about 100 fb$^{-1}$ is required for
the Higgs detection using this four charged lepton signature.
 
In this letter we focus exclusively on the 
hitherto difficult mass region $(ii)
$ with
\barr
155\rm{~GeV} < M_{H} < 180 \rm{~GeV}. \lab{massrange}
\earr
We show that despite the absence of a narrow mass peak the decay
\barr
H\ra W^{+}W^{-} \ra (\ell^{+}\nu)(\ell^{'-}{\bar\nu}),
\qquad \ell,\ell'=e,\mu, \tau(\rightarrow \ell \nu \bar{\nu}),
\lab{signal}
\earr
provides a straight forward discovery channel, especially in this 
mass range.

\section{ $\mathbf { H\ra W^{+}W^{-} \ra 
(\ell^{+}\nu)(\ell^{'-}{\bar\nu})}$ }

The Higgs decay to two W--bosons as well as the branching ratio, 
was first
calculated in Ref.\cite{rizzo} at tree-level. The one-loop result 
was obtained
shortly thereafter \cite{jeger}. In the mass range \eq{massrange} 
the Higgs
decay to two $W^\pm$ bosons is dominant with a branching ratio
close to
unity. For $W^\pm$ and $Z^0$ boson decays, isolated high $p_{t}$ 
electrons and
muons are typically clean and detectable signs of events. 
Despite the larger
branching ratios, the identification of $W^{\pm}$ and $Z^{0}$ 
using the decays
into jets is difficult to distinguish from the abundant jet 
background at the LHC even if one ($W,Z$) decays leptonically
\cite{qcdbkgd}. We therefore consider only the leptonic final 
states,
\barr
W^\pm &\ra& \ell^\pm \nu_\ell,\qquad \ell=e,\mu, \\
W^\pm &\ra & \tau^\pm \nu_\tau\,\ra\, \ell^\pm\nu_\ell\nu_\tau.  
\lab{wtau}
\earr
As a result, about 7\% of all $W$--pair events will have two 
oppositely charged
leptons $(e, \mu)$ \cite{pdg} and at least two neutrinos resulting 
in a
considerable missing transverse momentum imbalance. 
This should be compared to
the gold plated Higgs decay signature into two $Z$ bosons 
followed by their
leptonic decays. While the latter channel provides a narrow mass 
peak, it
suffers from the low $Z$ branching ratio to electron and muon pairs 
with a four charged lepton branching ratio of less 
than $0.45\%$ \cite{pdg}.
Combining this, with the factor of roughly 20--30 for the 
branching ratio of
the Higgs to $W^{+}W^{-}$ with respect to $Z^{0*}Z^{0}$ in 
the mass range
\eq{massrange} \cite{cms,atlas,lhczoltan,hunter}, we see that 
the decay signal
\eq{signal} is more than two orders of magnitude above 
the gold plated
signature. This large rate can therefore be expected to 
compensate for the
absence of a narrow mass peak due to the two undetected neutrinos.
 
For an intermediate mass Higgs, the fully leptonic decay 
signature \eq{signal}
was first studied by Glover \etal \cite{glover} for 
the LHC ($\sqrt{s}=16$ TeV). They explicitly did not consider the 
background from $t{\bar t}$ production and focused 
on the ``irreducible" 
background from $W^\pm$
pair production, for which they did not include any 
discriminating cuts. They
concluded that the final state \eq{signal} should provide 
detectable Higgs
signals at the LHC.
Subsequently, Barger \etal \cite{barger} performed a more 
detailed parton level
analysis of this signature for the LHC ($\sqrt{s}=16$ TeV). They
went beyond \cite{glover} to include the 
significant $t{\bar t}$ background for
$m_{top}=150,\,200$ GeV (the top quark mass was not yet known). 
Furthermore, in
their study, $W^{\pm}$'s from signal and background are 
simulated only with
their leptonic decays to e$\nu$ and $\mu \nu$. They did 
not include the mode
\eq{wtau}. Otherwise, their cuts are very similar to 
those in \cite{glover}, in
particular they also do not perform cuts to reduce 
the ``irreducible" $WW$
background. They again concluded, despite a much worse 
signal to background
ratio, that this channel should provide a reasonable possibility 
to detect the Higgs
in this mass range.
However, they also point out that 
a more detailed study
including hadronisation will eventually be necessary to 
substantiate the parton
level results.

Since then this Higgs signal has been ignored as a
discovery channel and the mass region between 155 GeV and 180 GeV 
has been identified as a problem 
area of the Higgs detection at the LHC
\cite{froidevaux_lhc,cms,atlas}. The motivation of this study is 
to demonstrate
that the potentially last Higgs-search gap at the LHC can be 
closed using the
specific decay \eq{signal}.

In the study described in the following, we go in many respects 
far beyond the
previous theoretical studies \cite{glover,barger}. We have 
included the decay
\eq{wtau} in the signal and in the background. Furthermore 
we have included the
background process\footnote{We are grateful to 
L. Rurua, who reminded us of this potential background.}
$gg\ra Wtb$, which is of the same order as top pair
production after the initial set of cuts. Most importantly, 
in all cases a full
simulation of QCD processes including hadronisation processes 
is done using the
PYTHIA Monte Carlo \cite{pythia} with the default   
CTEQ2L set of structure functions.
 
We find that the cuts 
previously employed
\cite{glover,barger} are then no longer sufficient. We have thus 
included new cuts in particular also 
to discriminate against the ``irreducible" $W^+W^-$ background.
Furthermore, the criteria used to select signal and background 
events are
chosen such that they can easily be fulfilled by the proposed 
ATLAS and CMS
experiments. As the proposed criteria are robust and relatively 
simple, they
are necessarily not optimised and significant improvements are 
possible once
the real detector behaviours are known. However, the proposed 
analysis
demonstrates the potential of a fast Higgs discovery in this 
mass range.

\section {Selection Cuts}
The goal of this analysis is to show that the expected large 
rate of the $pp
\ra H \ra W^{+}W^{-} \ra \ell^{+} \nu \ell^{'-} \bar{\nu}$ can 
compensate the
absence of a narrow mass peak. In order to perform a consistent 
analysis, the
PYTHIA Monte Carlo program is used for the simulation of 
signal and background
events. If not specified otherwise, the program is used with 
the default
parameter setting. For all production and decay processes we have 
considered
only the tree--level calculation as
the 1--loop calculations do not exist
for all relevant background processes. This allows a consistent 
analysis for signal and background and our results 
can easily be compared with previous studies using 
the four--charged--leptons 
channel. We thus do not include any K--factors. 
It is however worth pointing out that 
the 1--loop corrections
to the main Higgs production process via gluon--gluon fusion 
are large and
positive \cite{spira1}  
whereas the corrections to the 2--to--2 and
2--to--3 background processes are expected to be 
smaller \cite{spira2}.

The first set of selection criteria listed below selects 
relatively central
events with two isolated charged leptons (electrons or muons) 
with large
missing energy due to the two (or more) neutrinos. 
These cuts are mostly comparable to \cite{glover, barger}.
Starting from this initial set of cuts, one can concentrate on the 
differences between events 
originating from the Higgs production and from 
the so called ``irreducible"
background from continuum production of $pp \ra W^{+} W^{-} X$ 
events.
Even though the criteria strongly reduce
the events where the leptons originate from $Z$ decays, they are 
not yet sufficient for a complete removal of this possible 
background.
As will be shown below, this background is essentially removed 
by a stronger requirement, cut 9, on the opening angle $\phi$ 
between the two leptons in the plane transverse to the beam.
\begin{enumerate}
\item The event should contain two leptons with opposite charge 
each with a
minimal $p_{t}$ of 10 GeV. At least one of the two leptons should 
have a
$p_{t}$ of more than 20 GeV. Furthermore, the two leptons should 
be separated
in space by more than 10$^{\circ}$.
\item
The pseudorapidity $|\eta|$ of each lepton should be smaller 
than 2.
For events which contain additional isolated photons with a 
minimal $p_{t}$ of 20 GeV and $|\eta|$ smaller than 2.4 the 
photon 
four momentum vector is added to the dilepton system. 
\item
In order to have isolated leptons it is required that the 
energy sum
originating from hadrons and photons found within a cone of 
20 degree half
angle around the lepton direction should be smaller than 5 GeV.
\item
The dilepton mass, $m_{\ell \ell (\gamma)}$, has to be 
smaller than 80 GeV.
\item
The missing $p_{t}$ of the event, required to balance the 
$p_{t}$ vector sum of 
the two leptons, $\ell \ell (\gamma)$, should be larger 
than 20 GeV.
\item
The two leptons should not be back to back in the plane 
transverse to the beam
direction. The opening angle between the two leptons in 
this plane 
should be smaller than 135$^{\circ}$. In order to remove 
the backgrounds 
from $Z$ decays, this cut will be strengthened considerable 
in the following.
\item
Events which have a jet with a $p_{t}$ of more than 20 GeV 
and a pseudorapidity
$|\eta|$ of less than 2.4 are removed.
\end{enumerate}

Dilepton events, originating from the decays of $W$ and $Z$ bosons 
are selected
with criteria 1--3. Lepton pairs $\ell \ell (\gamma)$
originating from single $Z$ production
with subsequent decays to leptons, including the leptons coming 
from
decays of $\tau$ leptons are mostly removed with criteria 4--7. 
Backgrounds from $t\bar{t}$ and $Wtb$ production, \cite{wtb},
are reduced strongly by the jet veto, criterion 7.
It can be expected that even better background rejection factors 
can be obtained if some jet detection is possible up to larger 
values of $|\eta|$.
  
\begin{table}[t]
\vspace{0.3cm}
\begin{center}
\begin{tabular}{|c|c|c|c|c|}
\hline
\multicolumn{2}{|c|}{LHC 14 TeV}&\multicolumn{3}{|c|}
{Accepted event fraction} \\
\hline
reaction $pp \rightarrow X$ & $\sigma \times
BR^{2}$ [pb]
& cut 1-3 & cut 4-6 & cut 7 \\
\hline
$pp \rightarrow H \rightarrow W^{+}W^{-}$ ($m_{H}=170$ GeV)
&1.24  & 0.21 & 0.18 & 0.080\\
$pp \rightarrow W^{+}W^{-}$  &7.4  & 0.14 & 0.055& 0.039\\
$pp \rightarrow t \bar{t}$ ($m_{t}=175$ GeV)  & 62.0 & 0.17 &
0.070& 0.001\\
$pp \rightarrow W t b$ ($m_{t}=175$ GeV)      &  $\approx$6 & 0.17 &
0.092& 0.013 \\
\hline
$pp \rightarrow Z W \rightarrow \ell^{+} \ell^{-} \ell \nu $
& 0.86 & 0.23 & 0.054 & 0.026       \\
$pp \rightarrow Z Z \rightarrow$ 4--leptons & 1.05 & 0.13  & 
0.016& 0.007       \\
\hline
$pp \rightarrow Z \rightarrow \tau^{+} \tau^{-}$
& 1400 & 0.007  & 0.0004&  0.00009 \\
$pp \rightarrow Z \rightarrow e^{+}e^{-}, \mu^{+} \mu^{-}$
& 2800 & 0.22  & 0.0004&  0.00012 \\
\hline
\end{tabular}\vspace{0.3cm}
\end{center}
\caption{The expected signal and background event rates using 
the cross section
estimates with the CTEQ2L structure functions and with the first 
set of
selection criteria. In all cases only the leptonic $W$ branching 
ratios are
simulated ($W \rightarrow \ell^{\pm} \nu$ with $\ell^{\pm}$ 
being electrons,
muons or $\tau$). 
For the production of $ZZ$ events, the cross section 
is obtained including the $Z$ decays into charged leptons and 
neutrinos. 
For the production of $WZ$ and for the single $Z$ production 
only the
Z decays to charged leptons including the 
subsequent $\tau$ decays are simulated.}
\end{table}
The estimated cross sections before and after the above 
criteria for different
signal and background processes, 
including the leptonic branching ratios of the
$W$'s and $Z$ are given in table 1. As can be seen from table 1, 
the above
set of selection criteria reduces background from 
$pp \ra Z X$ events by a factor of about $10^{4}$. 
For the following 
we concentrate on the  
remaining background from continuum production
of $W^{+}W^{-}$, $t\bar{t}$ and $W t b$ events.

In order to distinguish a possible signal from the remaining 
background we use
the following criteria:
\begin{enumerate}
\setcounter{enumi}{7}
\item The polar angle $\theta_{\ell^{+} \ell^{-}}$ of the 
dilepton system,
reconstructed from the vector sum of their measured momenta 
should fulfill
$|\cos \theta_{\ell^{+} \ell^{-}} | < 0.8$.
\item The opening angle $\phi$
between the two charged leptons in the plane transverse
to the beam should be between 10$^{\circ}$--45$^{\circ}$.
\item
The mass, estimated for the assumed $W^{+}W^{-}$ system 
should be larger than
140 GeV. The mass is estimated from the approximation that 
the two neutrinos
compensate the $p_{t}$ of the two charged leptons, e.g. 
assuming  $p_{t}(H)
\approx 0$, and that the mass of the undetected 
neutrinos is on average
equal to the mass of the two charged leptons. The energy 
carried by the two
neutrinos is thus approximated 
with $E_{\nu \nu} = \sqrt{ m_{\ell \ell}^2 +
p_{t}^{2} (\ell \ell)}$. With this approximation a 
broad mass distribution,
with a mean value in agreement with the simulated 
Higgs mass and a large rms of
about 55 GeV, is obtained.
\item
The opening angle $\theta^{*}$ between the lepton with 
the larger $p_{t}$,
boosted to the dilepton rest frame and the momentum vector 
of the dilepton
system should fulfill the condition 
$ 0. <  \cos \theta^{*}_{\ell^{+} \ell^{-}}  < 0.3$.
\end{enumerate}

Condition eight exploits the smaller boost of the 
candidate events, originating
from the gluon--gluon fusion process. 
A large fraction of the continuum $W^{+}
W^{-}$ background originates from valence--quark 
sea--antiquark scattering with
a relatively large momentum imbalance, resulting in a 
boosted $W^{+}W^{-}$
system, as shown in figure 1.

Criterion nine makes use of the spin 
correlation between the $W^{+}W^{-}$ pair.
The potential discriminating power of this correlation 
in the Higgs search has
previously been pointed out by C. A. Nelson \cite{nelson}. 
$W$ pairs
originating from the decay of a scalar have to have opposite 
spin orientation.
Due the V--A structure in the W decay, the left handed $e^{-}$ 
(right handed
$e^{+}$) is emitted along the $W^{-}$ ($W^{+}$) spin. As a result, 
one of the
two charged leptons is emitted along the momentum direction of 
the two $W$'s
while the other one is emitted in the opposite direction. 
For the considered
Higgs mass range, a small opening angle between the two charged 
leptons can be
expected for signal events while the backgrounds will show an 
almost symmetric
distribution.
The discriminating power of this criterion is shown in figure 2.
As can be seen the leptons originating from Higgs decays have a
relatively small opening angle while the ones coming from 
continuum
$W^{+}W^{-}$ and $t \bar{t}$ events show essentially a symmetric 
distribution.

The estimated invariant mass of the
$\ell^{+}\ell^{-} \nu \nu$ system, shown
in figure 3, is unfortunately very broad. Nevertheless
it discriminates to some extent between signal and background.
\begin{figure}[htb]
\begin{center}\mbox{
\epsfig{file=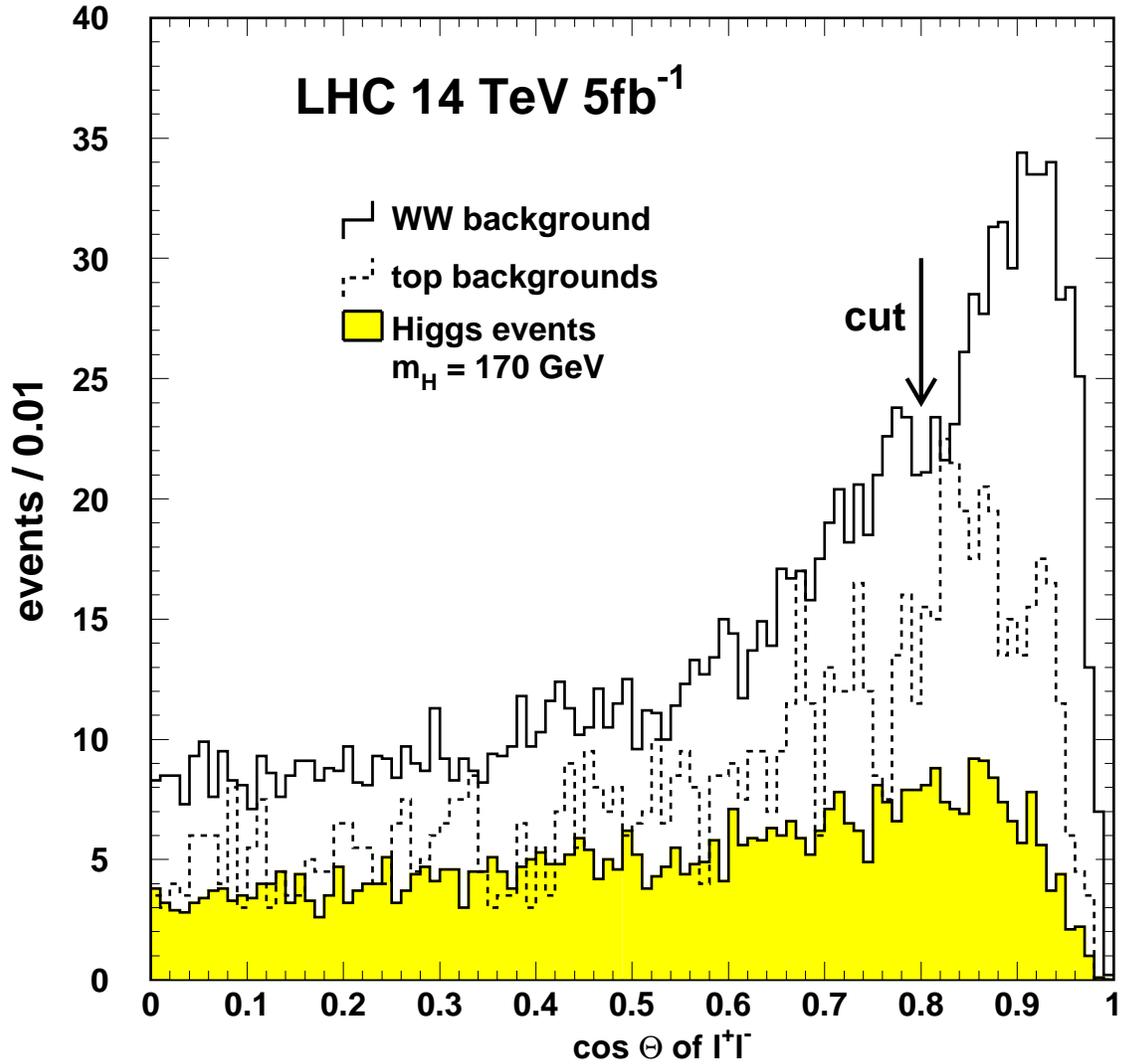,
height=16 cm,width=16cm}}
\end{center}
\caption[fig1]{$| \cos \theta |$ 
distribution of the dilepton  
system with respect to the beam direction 
for Higgs signal 
and background events. 
}
\end{figure}

\begin{figure}[htb]
\begin{center}\mbox{
\epsfig{file=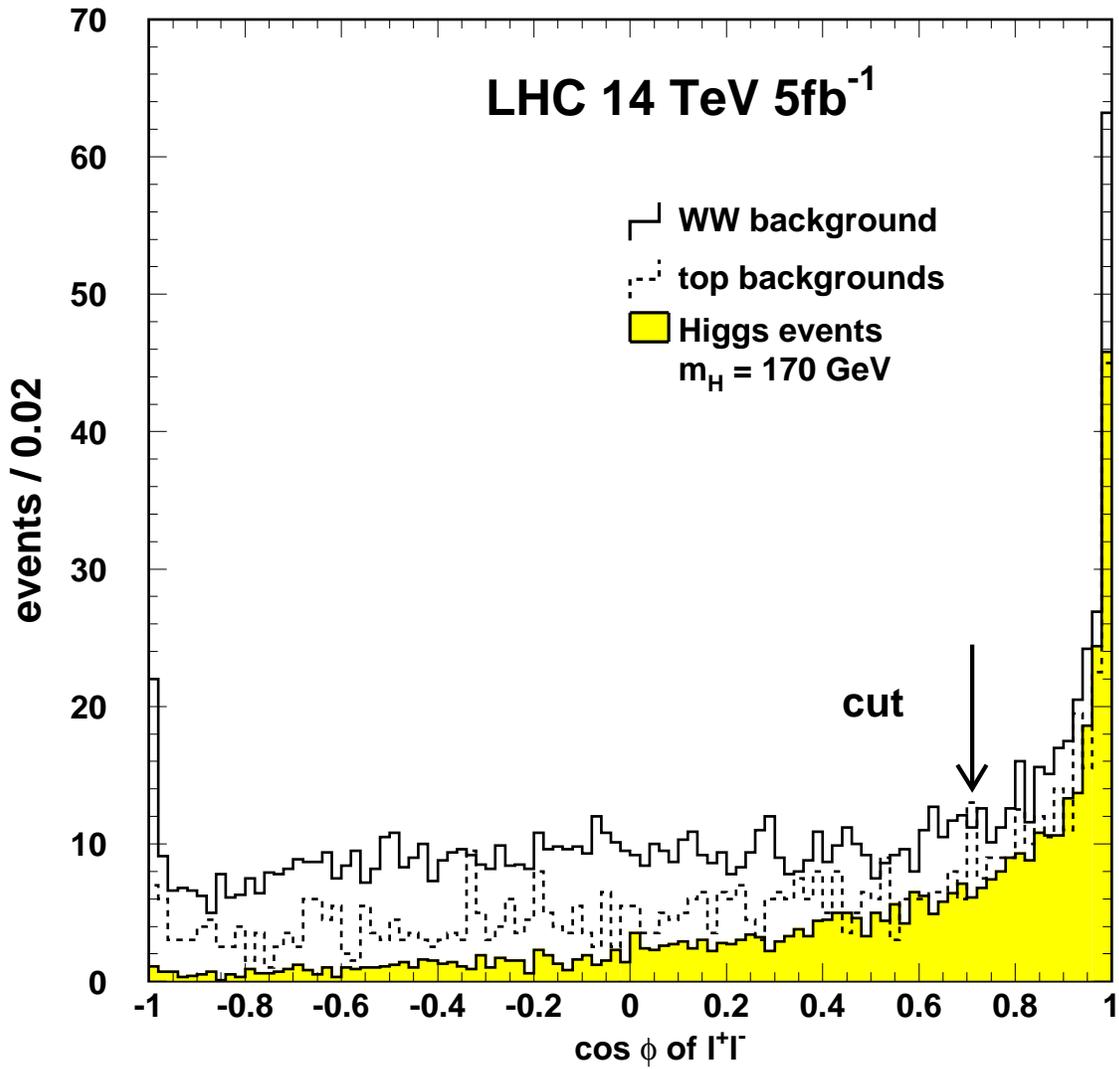,
height=16 cm,width=16cm}}
\end{center}
\caption[fig2]{$ \cos \phi $ 
distribution of the dilepton  
system in the plane transverse to the beam direction for Higgs 
signal 
and background events, 
cut number 6 has not yet been applied.
}
\end{figure}
\clearpage

\begin{figure}[htb]
\begin{center}\mbox{
\epsfig{file=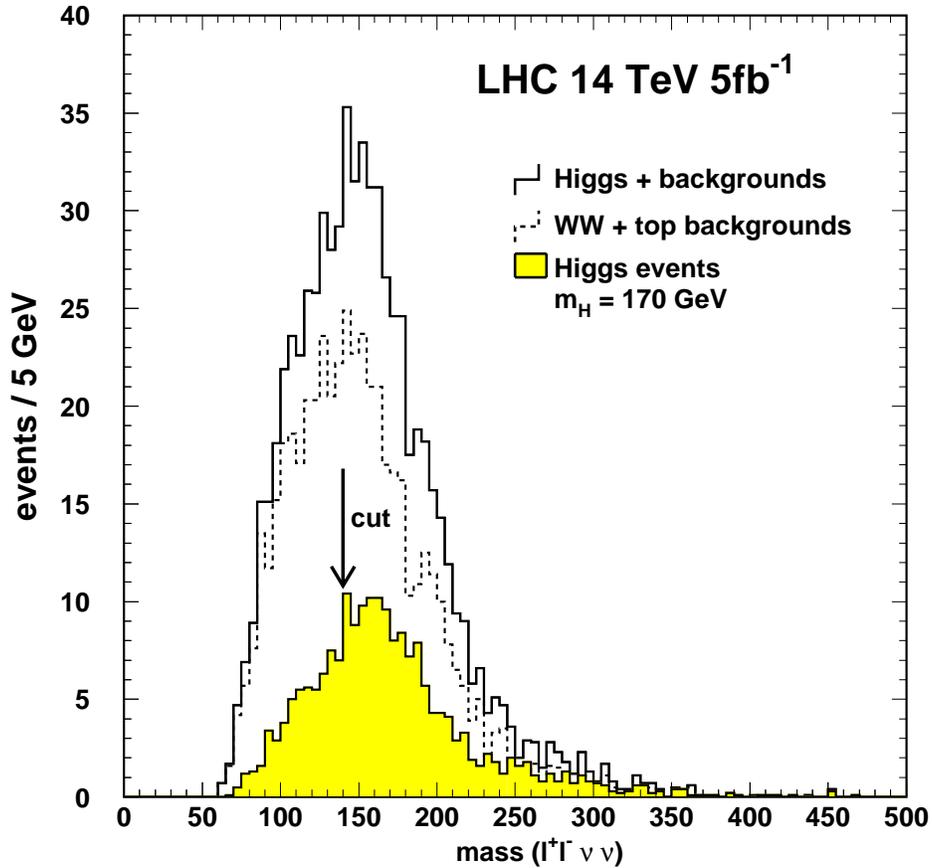,
height=13 cm,width=13cm}}
\end{center}

\caption[fig2]{Estimated invariant mass of the 
$\ell \ell \nu \nu $ system for Higgs signal 
and background events. 
}
\end{figure}

Finally, the two charged leptons from Higgs events
show a smaller momentum spread than the ones from
the background. This fact shows up nicely in figure 4,
where the signal events show a peak like structure
for small values of $\cos \theta^{*}_{\ell^{+} \ell^{-}}$
while the backgrounds show a strong increase
to larger values. This distribution has also
some sensitivity to the Higgs mass and can therefore be used
for a mass estimate.

\begin{table}[t]
\vspace{0.3cm}
\begin{center}
\begin{tabular}{|c|c|c|c|c|c|}
\hline
\multicolumn{6}{|c|}{Structure Function CTEQ2L} \\
\hline
\multicolumn{2}{|c|}{LHC 14 TeV}&\multicolumn{4}{|c|}
{Expected event rate for 5 fb$^{-1}$} \\
\hline
reaction $pp \rightarrow X$ & $\sigma \times
Br^{2}$ [pb]
& cut 1--7 & cut 8--9 & cut 10 & cut 11 \\
\hline
$pp\rightarrow H $ ($m_{H}=155 $GeV) & 1.09& 426& 168 & 99& 49\\
$pp\rightarrow H $ ($m_{H}=160 $GeV) & 1.25& 508& 212 & 140& 78\\
$pp\rightarrow H $ ($m_{H}=165 $GeV) & 1.27& 520& 220 & 151& 86\\
$pp\rightarrow H $ ($m_{H}=170 $GeV) & 1.24& 497& 201 & 147& 74\\
$pp\rightarrow H $ ($m_{H}=175 $GeV) & 1.19& 462& 176 & 129& 59\\
$pp\rightarrow H $ ($m_{H}=180 $GeV) & 1.11& 398& 151 & 112& 47\\
\hline
$pp \rightarrow W^{+}W^{-}$  & 7.4& 1458 & 273 & 130& 38\\
$pp \rightarrow t \bar{t}$ ($m_{t}=175 $GeV)  & 62.5& 441 & 104 &
72& 18 \\
$pp \rightarrow W t b$ ($m_{t}=175$ GeV) & $\approx 6$ & 397 & 
110 & 70& 24\\
\hline
$pp \rightarrow ZZ, WZ$ & 1.9 & 150 & 31 & 16 & 5\\
$pp \rightarrow Z $ & 4200 & 2355 & 49 & 24 & 7 ($\leq$ 13) \\
\hline
$\sum$ all backgrounds   & -- & 4781 & 567 & 312 & 92\\
\hline
\multicolumn{6}{|c|}{Structure Function EHLQ set 2} \\
\hline
$pp \rightarrow H $ ($m_{H}=170$ GeV) & 1.7 & 653 & 263 & 
185& 92\\
$pp \rightarrow W^{+}W^{-}$  & 5.9& 1152 & 231 & 110& 35\\
$pp \rightarrow t \bar{t}$ ($m_{t}=175$ GeV)  & 95& 741 & 
163 & 104& 24\\
\hline

\end{tabular}\vspace{0.3cm}
\end{center}
\caption{The expected event rates
for signal and background for an integrated luminosity
of 5 fb$^{-1}$ using a PYTHIA simulation and CTEQ2L.
For a comparison the most important background rates
are also given for the EHLQ structure functions.}
\end{table}

Table 2 shows the number of accepted signal and background
events for an integrated luminosity of 5 fb$^{-1}$ at the
LHC with 14 TeV center of mass energy.
Taking the signal and background event rates for the considered
luminosity of 5 fb$^{-1}$ statistical significant signals
appear already after cuts 1--7. However, as signal
and background cross sections are not well known,
a signal to background ratio of about 1 to $\approx$10 
is perhaps not sufficient.
As has been discussed above, with the
subsequent criteria, signal to background ratios
of about 1 to 1 can be obtained while keeping sizeable signal rates.
Furthermore, as can be seen from figures 1--4, 
absolute background rates can
be estimated from several distributions where clear separations
between a Higgs signal and backgrounds are obtainable.

Using all criteria we obtain
Higgs signals with a significance between 5--10 sigma
for the considered mass range and an integrated luminosity
of about 5 fb$^{-1}$.
This result should be compared to the significance
obtained for the gold plated four charged lepton channel
where about 100 fb$^{-1}$ are required for a 5 sigma
signal.

\begin{figure}[htb]
\begin{center}\mbox{
\epsfig{file=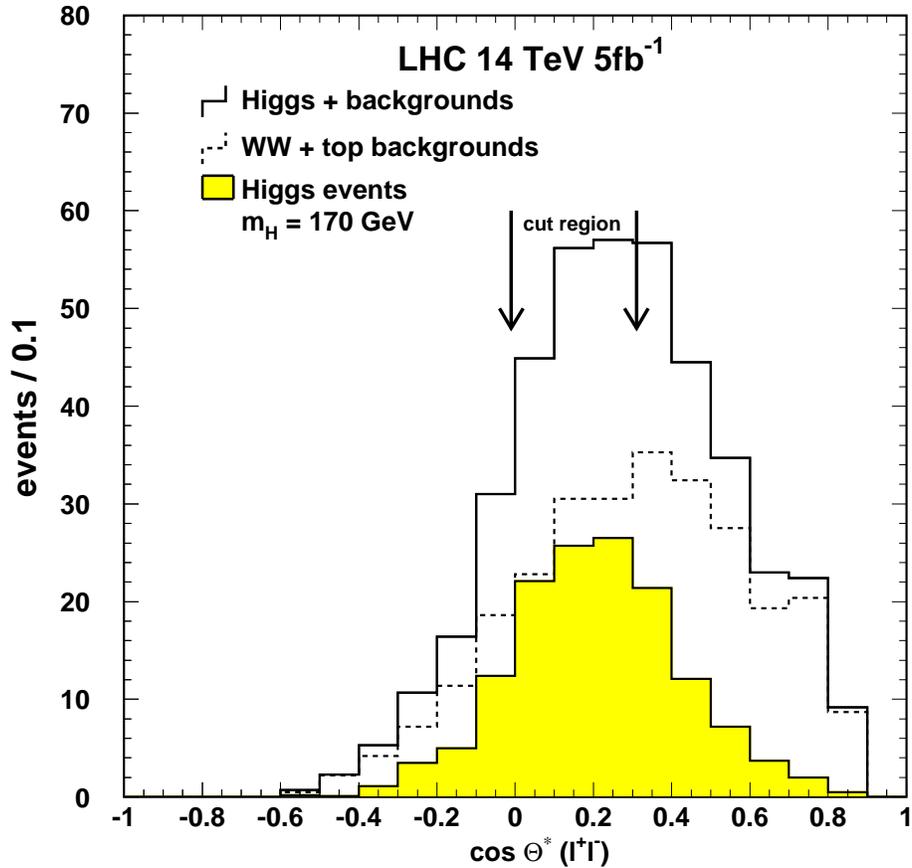,
height=13 cm,width=13cm}}
\end{center}

\caption[fig4]{$\cos \theta^{*}$ distribution in the dilepton 
rest frame for Higgs signal 
and background events. 
}
\end{figure}
\section{Conclusions} 
We have reconsidered the signature of two leptons plus missing 
energy
as a signal for the Higgs boson decay mode $H\ra W^+W^-$.
It is found that the Higgs detection in the
previously considered difficult mass range
between 155--180 GeV appears to be relatively easy for this
decay signature. Using a few simple experimental
criteria, clear differences between
signal and backgrounds are obtained allowing a 5--10 sigma
Higgs signal detection with an integrated luminosity of 
about 5 fb$^{-1}$.
We thus conclude that for the considered mass range
events with two leptons plus missing energy will provide, 
the Higgs discovery signature at the LHC.

\vspace{2.cm}
{\bf \large Acknowledgements}

{\small
We would like to thank
U. Baur and M. Spira for discussions. We are also 
grateful to T. Sj\"ostrand for his immediate help on how to 
use PYTHIA in the most efficient way
for the $Wtb$ background.}

\end{document}